\documentclass{article}
\usepackage{spconf,amsmath,graphicx}
\usepackage{subfigure}
\usepackage{amssymb}
\usepackage{multirow}
\usepackage{booktabs}
\usepackage{cite}
\usepackage{comment}
\usepackage{pifont}
\def\tabf{{\ \ }}
\title{FunCodec: A Fundamental, Reproducible and Integrable Open-source Toolkit for Neural Speech Codec}
\name{Zhihao Du, Shiliang Zhang, Kai Hu, Siqi Zheng\thanks{Demos are available at: https://funcodec.github.io}}
\address{Speech Lab of DAMO Academy, Alibaba Group, China\\
\texttt{\{neo.dzh,sly.zsl\}@alibaba-inc.com}}
\begin{document}
%\ninept
%
\maketitle
%

% set 9pt font size
\ninept
\begin{abstract}
This paper presents FunCodec, a fundamental neural speech codec toolkit, which is an extension of the open-source speech processing toolkit FunASR.
FunCodec provides reproducible training recipes and inference scripts for the latest neural speech codec models, such as SoundStream and Encodec.
Thanks to the unified design with FunASR, FunCodec can be easily integrated into downstream tasks, such as speech recognition.
Along with FunCodec, pre-trained models are also provided, which can be used for academic or generalized purposes.
Based on the toolkit, we further propose the frequency-domain codec models, FreqCodec, which can achieve comparable speech quality with much lower computation and parameter complexity.
% This paper describes the design of FunCodec and its features different from other toolkits. 
Experimental results show that, under the same compression ratio, FunCodec can achieve better reconstruction quality compared with other toolkits and released models.
We also demonstrate that the pre-trained models are suitable for downstream tasks, including automatic speech recognition and personalized text-to-speech synthesis.
This toolkit is publicly available at \texttt{https://github.com/alibaba-damo-academy/ FunCodec}.
% 1. implement soundstream and encodec, integrated to downstream tasks
% 2. release pre-treind models and better than others
% 3. we propose FreqCodec.
\end{abstract}
\section{Introduction}
\label{sec:intro}

Speech codecs are designed to compress and decompress speech signals for efficient transmission and storage.
% They consist of an encoder to encode speech into a compact representation and a decoder to reconstruct the signal.
They consist of an encoder, which encodes speech into a compact representation, and a decoder to reconstruct the signal.
Traditional speech codecs rely on a carefully designed pipeline that incorporates expert knowledge of psycho-acoustics and speech synthesis to achieve efficient coding \cite{DBLP:journals/rfc/rfc6716, DBLP:conf/icassp/DietzMEMNP0WLVK15}.

Thanks to advancements in deep learning techniques, neural speech codecs have been introduced, demonstrating superior performance compared to traditional speech codecs.
In neural codec models, raw waveforms are fed into deep neural network-based encoders to extract compact representations. This is followed by a residual vector quantizer (RVQ) \cite{DBLP:conf/nips/KumarKBGTSBBC19, DBLP:journals/taslp/ZeghidourLOST22} to obtain parallel token streams. 
Meanwhile, a neural network-based decoder is also trained alongside the encoder and RVQ to reconstruct the signal.
Building upon the progress in text-to-speech synthesis \cite{DBLP:conf/nips/KongKB20}, adversarial training losses are employed to enhance reconstruction quality.
% , in which multiple discriminators are employed to identify the original signals and reconstructed ones.
There are two popular neural codec models, SounStream \cite{DBLP:journals/taslp/ZeghidourLOST22} and Encodec \cite{defossez2022highfi}.
While SoundStream utilizes streaming SEANet \cite{DBLP:conf/interspeech/TagliasacchiLMR20, DBLP:conf/icassp/LiTRUR21} as its encoder and decoder, Encodec incorporates extra LSTM \cite{DBLP:journals/neco/HochreiterS97} layers and a Transformer-based language model \cite{DBLP:conf/nips/VaswaniSPUJGKP17} to improve the sequence modeling ability. Following this line of work, extensive efforts have been made on reducing the bit rate \cite{DBLP:journals/icassp/low-bitrate, DBLP:journals/icassp/lmcodec, DBLP:journals/icassp/Neural}. 
Additionally, the modified discrete cosine transform (MDCT) domain has also been explored \cite{DBLP:journals/icassp/MDCTNet,DBLP:journals/icassp/MDCTdomain}.

\begin{figure}[t!]
	\centering
	\includegraphics[width=1.0\linewidth]{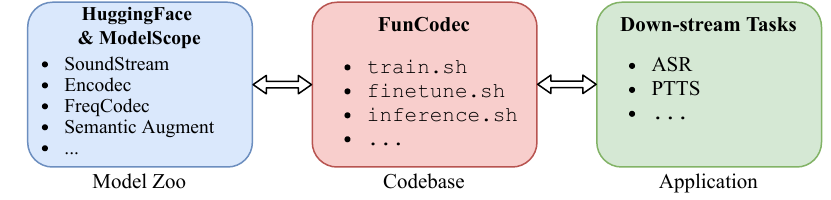}
	\vspace{-0.5cm}
	\caption{Overview of FunCodec design.}
	\vspace{-0.5cm}
	\label{fig:design}
\end{figure}

Although neural speech codecs were originally proposed to compress signals for telecommunication, they can also be used to extract discrete speech representations in generative models. In VALL-E \cite{DBLP:journals/corr/abs-2301-02111}, text and speech tokens are joined into a sequence, and a language model is trained to estimate their probabilities. This formulation demonstrates impressive zero-shot synthesis capability. 
Moreover, neural speech codecs facilitate the modeling of speech and text within a single framework, enabling the model to both listen and speak \cite{DBLP:journals/corr/abs-2305-16107,DBLP:journals/corr/abs-2306-12925}. 
% In these frameworks, the speech and text related tasks can be collectively regarded as sequence generation problems by joining the inputs and outputs with special task-related tokens, such speech recognition, text translation, text-to-speech synthesis and et al.
% the speech codec model is treated as a tokenizer,  is represented as discrete tokens, which can be unified modeling with text tokens. As a result, an unified model can be trained to perform many text/speech-to-speech/text tasks, such as, speech recognition, speech/text translation, text-to-speech synthesis and et al.
Recently, several speech codec toolkits have been released for telecommunications purposes \cite{10096509,desc-dac}. However, there is still a lack of open-source toolkits that provide a reproducible and integrable framework for developing and evaluating neural speech codecs in the context of speech-text modeling.

To address this gap, we present FunCodec, a fundamental, reproducible, and integrable open-source toolkit for neural speech codecs. 
FunCodec provides a versatile platform enabling researchers to build, train, and evaluate various neural speech codecs.
Fig.\ref{fig:design} shows an overview of the FunCodec design. 
% As shown in Fig.\ref{fig:design}, FunCodec has several unique features:
The contributions of FunCodec are as follows: 
(1) The open-source codebase provides recipes to finetune pre-trained models or train a model from scratch. 
(2) Frequency-domain codec (FreqCodec) models are proposed, which can achieve comparable performance with less parameters and lower computation complexity. 
(3) The impact of semantic information is evaluated for speech codec, which improves the speech quality under low bit rate.
(4) Pre-trained academic and generalized models are released through Huggingface and ModelScope \footnote{https://www.modelscope.cn/models?page=1\&tasks=audio-codec\&type=audio}.
(5) Inference and evaluation scripts are also provided, which support batch mode to fully utilize the parallelism capability of GPUs. 

% The open-source codebase provides recipes to finetune pre-trained models or train a model from scratch.
% Inference and evaluation scripts are also provided, which support batch mode to fully utilize the parallelism ability of GPUs.

\begin{table}[t!] 
	\centering 
	\caption{Feature comparison of FunCodec and other open-source speech codec toolkits.} 
	\vspace{0.1cm}
	\label{tab:compare} 
	\setlength\tabcolsep{1.0pt}
	\begin{tabular}{l|c | c | c | c | c} 
		\toprule 
		& Encodec & EncTrain & {Dac.} & {AudioDec} &  {FunCodec}\\
		\midrule
		{Organization} & Facebook & Mikxox & Descript & Facebook & Alibaba \\
		{Released models} & 2 & 1 & 3 & 3 & 7 \\ % 2 academic models, 2 industrial models, 2 freq models, 1 ppg model
		{Training recipe} & \ding{55} & \ding{51} & \ding{51} & \ding{51} & \ding{51} \\
		{Training stages} & \ding{55} & 1 & 1 & 2 & 1 \\
		% {Supported Disc.} & \ding{55} & MSD & MSD,MPD,MSPD & MSD,MPD & MSD,MPD,MSPD,MFD \\
		{Discriminators} & \ding{55} & 1 & 3 & 2 & 4 \\
		{Dist. training} & \ding{55} & \ding{55} & \ding{51} & \ding{51} & \ding{51} \\
		{K-means init.} & \ding{55} & \ding{51} & \ding{55} & \ding{55} & \ding{51} \\
		{Low-frame-ratio} & \ding{55} & \ding{55} & \ding{55} & \ding{55} & \ding{51} \\
		{Freq. domain} & \ding{55} & \ding{55} & \ding{55} & \ding{55} & \ding{51} \\
		{Semantic aug.} & \ding{55} & \ding{55} & \ding{55} & \ding{55} & \ding{51} \\
		\bottomrule
	\end{tabular}
\vspace{-0.4cm}
\end{table}

\vspace{-0.1cm}
\section{Related Work}
\vspace{-0.1cm}
In this section, we compare FunCodec with other toolkits.  
We have chosen the following four toolkits available on Github:
\begin{itemize}
\item[$\bullet$] Encodec: \texttt{facebookresearch/encodec} \cite{defossez2022highfi},
\item[$\bullet$] EncTrainer: \texttt{Mikxox/EnCodec\_Trainer} \cite{EncTrainer},
\item[$\bullet$] Dac.: \texttt{descriptinc/descript-audio-codec} \cite{desc-dac},
\item[$\bullet$] AudioDec: \texttt{facebookresearch/AudioDec} \cite{10096509}.
\end{itemize}

Table \ref{tab:compare} summaries the differences between FunCodec and these toolkits. 
While the other toolkits offer a limited number of
pre-trained models, FunCodec provides seven models for both academic and generalized purposes.
This allows researchers to use them as a baseline system and also enables general users to directly apply them to downstream tasks.
Additionally, FunCodec provides comprehensive and efficient recipes that require only a single training stage.
To enhance the speech quality, FunCodec incorporates various discriminators, 
including multiple scale discriminator (MSD) \cite{DBLP:conf/nips/KumarKBGTSBBC19,DBLP:conf/interspeech/TagliasacchiLMR20}, multiple period discriminator (MPD) \cite{DBLP:conf/nips/KongKB20}, multiple short-time Fourier transformation discriminator (MSTFTD) \cite{DBLP:journals/taslp/ZeghidourLOST22}, and their combinations.
For training efficiency, FunCodec supports distributed training across multiple GPUs. 
Moreover, FunCodec ensures high inference efficiency by simultaneously producing token streams for all samples in a mini-batch.
% While other toolkits obtain tokens sample by sample, FunCodec can generate them for all samples in a mini-batch, resulting in much higher inference efficiency.
Furthermore, FunCodec enables k-means initialization for quantization codebooks, improving the code utilization \cite{DBLP:conf/nips/OordVK17,DBLP:conf/icassp/RenLHZCYZ22}.
Based on these features, FunCodec introduces low-frame-rate models. The frequency-domain transformation and semantic augmentation are also explored in FunCodec.

\section{FunCodec Toolkit}
FunCodec codebase consists of two main components: a library of neural network models and recipes for replicating the experiments.
The library part is written in python with PyTorch \cite{paszke2017automatic}. 
The recipes are all-in-one Bash scripts written in the Kaldi-style \cite{povey2011kaldi}.

\subsection{Model architecture}
\begin{figure}[t!]
	\centering
	\includegraphics[width=1.0\linewidth]{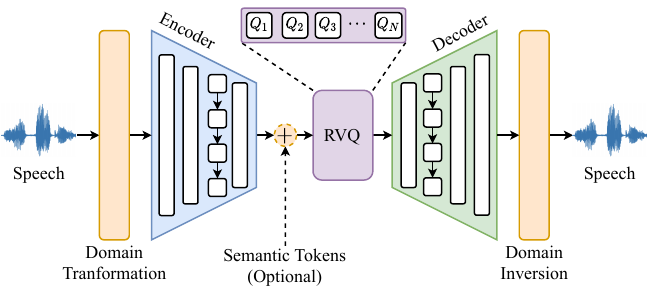}
	\vspace{-0.5cm}
	\caption{The overall architecture of the FunCodec models.}
	\vspace{-0.5cm}
	\label{fig:framework}
\end{figure}

The architecture of FunCodec models is depicted in Fig. \ref{fig:framework}.
Given a speech signal $x$, it is passed through the domain transformation module. For time-domain models like SoundStream and Encodec, the module functions as an identity mapping. However, for frequency-domain models, two representations $X_{\text{mag,ang}}$ and $X_{\text{mag,pha}}$ are explored:
\begin{equation}
\begin{split}
X &= \text{STFT}(x) \\
X_{\text{mag,ang}} &= \log\left(|X|\right), \text{angle}(X_i, X_r) \\ 
X_{\text{mag,pha}} &= \log\left(|X|\right), \frac{X_r}{|X|}, \frac{X_i}{|X|}
\end{split}
\end{equation}
where $X_r, X_i$ denote the real and imaginary parts of complex spectrum, respectively. $|\cdot|$ represents the norm of a complex value.
After the domain transformation module, the speech is inputted into an encoder to extract acoustic representations: $V_a=\text{Encoder}(X)$.
For time-domain models, we adopt the same SEANet architectures as Encodec and SoundStream.
In the case of frequency-domain models (FreqCodec), the encoder details are given in Table \ref{tab:model-details}. 
The decoder has a mirror architecture of the encoder.
% but due to page limitations, we omit its description here. 
More details can be found in our released codebase. Finally, a domain inversion module is utilized to reconstruct the raw waveforms from decoder outputs.
\begin{table}[htb]
	\renewcommand{\arraystretch}{1.2} 
	\vspace{-0.3cm}
	\caption{The encoder architecture details of FreqCodec. $S_b$ and $C_b$ denote the stride and filters of EncBlock $b$.}
	\vspace{0.1cm}
	\label{tab:model-details}
	\centering
	\setlength\tabcolsep{3pt}
	\footnotesize
	\begin{tabular}{| l | c | c | c |}
		\hline
		Layer Name & Inputs & Kernel, Stride & Outputs \\
		\hline
		Domain Trans. & $(1, t)$ & 512, 160 & $(*, T, 257)$ \\
		\hline
		% Encoder & $(3, T, 257)$ & -, $S$ & $(T/S, D)$ \\
		% \hline
		PreConv2D & $(*, T, 257)$ & $(7, 7),(1, 1)$ & $(C, T, 256)$ \\ 
		\hline
		EncBlock $\times B$ & $(C_b, T_b, F_b)$ & -, $(S_b,4)$ & $(2C_b, T_b/S_b, F_b/4 )$\\
		\hline
		\tabf Conv2D\_1 & $(C_b, T_b, F_b)$ & $(3, 3)$, $(1,1)$ & $(C_b/2, T_b, F_b)$ \\
		\tabf Conv2D\_2 & $(C_b/2, T_b, F_b)$ & $(1, 1)$, $(1,1)$ & $(C_b, T_b, F_b)$ \\
		\tabf Conv2D\_ds & $(C_b, T_b, F_b)$ & $(2S_b, 8)$, $(S_b,4)$ & $(2C_b, T_b/S_b, F_b/4)$ \\
		\hline
		Reshape & $(2^4C, T_B, 1)$ & - & $(T_B, 2^4C)$\\
		\hline
		LSTM & $(T_B, 2^4C)$ & $(2^4C, 2^4C)$ & $(T_B, 2^4C)$\\
		\hline
		OutLinear & $(T_B, 2^4C)$ & $(2^4C, D)$ & $(T_B, D)$\\
		\hline
	\end{tabular}
\end{table}
\vspace{-0.3cm}

\subsection{Semantic-augmented residual vector quantization}
To obtain discrete speech tokens, we employ a residual vector quantization (RVQ) module consisting of several quantizers:
\begin{equation}
	Q_n=\text{VQ}\left(Q_0-\sum_{i=1}^{n-1}{Q_i}\right)
\end{equation}
where $Q_n$ represents the outputs of $n$-th vector quantizer (VQ) and $Q_0$ represents the input of RVQ.
To improve code utilization, we initialize the VQ codebook by clustering the samples in the first mini-batch with k-means.
The codes are then updated using a moving average with a decay rate of $0.99$. 
Moreover, if a code is activated fewer than two times in a mini-batch, it will be reassigned.

In addition to the encoder outputs, we explore three methods to incorporate semantic information into the codec models:
\begin{equation}
\begin{split}
f_{\text{cat}}(V_a,V_s)&=\text{Concat}(\text{RVQ}(V_a),V_s) \\
f_{\text{add}}(V_a,V_s)&=\text{RVQ}(V_a)+V_s \\
f_{\text{res}}(V_a,V_s)&=\text{RVQ}(V_a-V_s)+V_s
\end{split}
\end{equation}
where $V_s$ denotes semantic tokens, such as frame-aligned phoneme labels and Hubert embeddings \cite{DBLP:journals/taslp/HsuBTLSM21}.
To make a single model operate across variable bitrates, structured quantization dropout is also implemented in FunCodec.
% Subsequently, a residual vector quantization (RVQ) module is used to obtain discrete speech tokens:

\subsection{Adversarial training objective with multiple discriminators}
The training objective consists of three components: reconstruction loss terms, adversarial loss terms, and the RVQ commit losses.
The L1 distance between original $x$ and reconstructed speech $\hat{x}$ is minimized over time domain: $\mathcal{L}_t(x,\hat{x})=||x-\hat{x}||_1$.
For the frequency domain, both L1 and L2 distances are minimized on multiple Mel and magnitude spectra:
\begin{equation}
\begin{split}
	\mathcal{L}_f(x, \hat{x})&=\frac{1}{|\alpha|}\sum_{i\in \alpha}({||\mathcal{S}_i(x)-\mathcal{S}_i(\hat{x})||_1 + ||\mathcal{S}_i(x)-\mathcal{S}_i(\hat{x})||_2 } \\
	& + ||\mathcal{M}_i(x)-\mathcal{M}_i(\hat{x})||_1 + ||\mathcal{M}_i(x)-\mathcal{M}_i(\hat{x})||_2)
\end{split}
\end{equation}
where, $\mathcal{S}_i$ and $\mathcal{M}_i$ represent the log-compressed power and Mel spectra with a window size of $2^{i}$ and a shift length of $2^{i}/4$. $\alpha$ is set to $[5,6,\dots,11]$. It worth noting that the log-compressed power spectrum loss improves speech quality in the middle and high frequencies, which is missed in other toolkits and models.

For adversarial losses, FunCodec incorporates several discriminators, including multi-scale discriminator (MSD), multi-period discriminator (MPD), multi-scale STFT-based (MSTFTD) discriminator. By providing a unified interface, FunCodec allows for various combinations of these discriminators, resulting in a more powerful discriminative ability. In addition, a ``feature'' matching loss is also involved:
\begin{equation}
\begin{gathered}
\mathcal{L}_{\text{adv}}(\hat{x})=\mathbb{E}_{\hat{x}}\left[\frac{1}{K}\sum_{k,t}{\frac{1}{T_k}{\max(0,1-\mathcal{D}_{k,t}(\hat{x}))}}\right] \\
\mathcal{L}_{\text{feat}}(x, \hat{x})=\mathbb{E}_{x,\hat{x}}\left[\frac{1}{KL}\sum_{k,t,l}{\frac{1}{T_k}{\left|\left|\mathcal{D}^{(l)}_{k,t}(x)-\mathcal{D}^{(l)}_{k,t}(\hat{x})\right|\right|_1}}\right] \\
\end{gathered}
\end{equation}
where $\mathcal{D}_{k,t}$ represents the output of discriminator $k$ at timestep $t$, and $\mathcal{D}^{(l)}_{k,t}$ denotes the outputs of layer $l$.

The commit loss takes into account the quantization errors of the whole RVQ module and sub-quantizers:
\begin{equation}
\mathcal{L}_\text{cm} = \left|\left| V-\text{RVQ}(V) \right|\right|_2 + \frac{1}{N}\sum_{i=1}^{N}{\left|\left|Q_{i-1}-\text{VQ}_i(Q_{i-1})\right|\right|_2}
\end{equation} 
where $V$ represents the inputs of RVQ, which can vary depending on the semantic augmentation methods used. The total training objective is obtained by summing up the individual loss terms:
\begin{equation}
\mathcal{L}=\lambda_{t}\mathcal{L}_{t}+\lambda_{f}\mathcal{L}_{f} + \lambda_{\text{adv}}\mathcal{L}_{\text{adv}} + \lambda_{\text{feat}}\mathcal{L}_{\text{feat}} + \lambda_{\text{cm}}\mathcal{L}_\text{cm}
\end{equation}

\section{Experimental Settings}
\subsection{Experimental condition}
We conduct experimental evaluations under two conditions, one for academic usage and the other for generalized purposes.
For academic usage, we employ the commonly-used LibriTTS corpus \cite{zen2019libritts} to train and evaluate models. This corpus consists of 585 hours of English speech.
We follow the official data partition of training, validation and test sets. 
For generalized usage, we train the model on a large-scale in-house dataset, which contains 27.68 million bilingual (English and Mandarin) utterances with a total duration of approximately 25,000 hours. 
To assess the generalization ability of our generalized model, we evaluate it on the test sets of multiple open-source corpora, including Librispeech \cite{panayotov2015librispeech}, aishell-1 \cite{bu2017aishell}, aishell-2 \cite{du2018aishell}, Wenet \cite{zhang2022wenetspeech} and Gigaspeech \cite{chen2021gigaspeech}.
In this paper, we resample all utterances to a sampling rate of 16k Hz.
It is worth noting that FunCodec also supports other commonly-used sampling rate, including 8k and 24k, among others.

We adopt the Virtual Speech Quality Objective Listener score (ViSQOL) \cite{hines2015visqol} as the primary evaluation metric, which ranges from 1 to 5, with higher scores indicating better quality.
To ensure a fair comparison, we introduce a measurement called token ratio (TKR), which indicates the number of tokens the model requires to represent one second of speech at a sampling rate of 16k.
Under the same TKR, models with higher ViSQOL scores are preferred.
% For the down-stream ASR task, the word error rate (WER) is employed to evaluate the recognition performance, which suggests the quantity of semantic information contained in the quantized codes.

\vspace{-0.3cm}
\subsection{Training details}
During the training stage, we randomly clip a continuous segment of 3.2 seconds from an utterance, which is considered as a training sample. 
Before being fed into the encoder, the segment undergoes root-mean-square (RMS) normalization. The reconstructed output is rescaled using inverse normalization to calculate losses.
% We perform root-mean-square (RMS) normalization on the segment before it is fed into the generator, and the reconstructed output is re-scaled with the inverse normalization to calculate losses. 
For the LibriTTS corpus, we train the models on two Tesla-V100 GPUs with a total batch size of 32. 
On the other hand, the generalized models are trained on four Tesla-A100 GPUs with a total batch size of 128.
% while the generalized models are trained on four Tesla-A100 GPUs with the total batch size of 128.
Under the adversarial training framework, we update the codec model 300,000 times.
To prevent the discriminator from becoming too dominant, we only update it when its loss exceeds that of the codec model.

The hyperparameters $\lambda_{t}$, $\lambda_{f}$, $\lambda_{\text{adv}}$, $\lambda_{\text{feat}}$ and $\lambda_{\text{cm}}$ are set to $1.0$, $1.0$, $\frac{1}{9}$, $\frac{100}{9}$ and $1.0$, respectively.
We find that this hyperparameter setting is suitable for speech signals, and the loss balancer \cite{defossez2022highfi} is also a work in progress.
For FreqCodec models, the speech segment is first transformed into a spectrogram using STFT with a window size of 512 and a shift size of 160.
% For the ASR models, we employ the 

\vspace{-0.3cm}
\section{Experimental Results}
\subsection{Evaluation of academic models on LibriTTS corpus}
\begin{table}[t!]
	\caption{Comparison of academic models in terms of ViSQOL scores on LibriTTS dataset. $\dagger$ means the model is causal.}
	\vspace{0.1cm}
	\label{tab:librtts}
	\centering
	\setlength\tabcolsep{3.5pt}
	\begin{tabular}{l | c | c c c c}
		\toprule
		Models & Stride & 400 TKR & 200 TKR & 100 TKR & 50 TKR \\
		\midrule
		SoundStream$^\dagger$ & 320 & 4.23 & 4.00 & 3.60 & 3.12\\
		SoundStream & 320 & 4.28 & 4.06 & 3.76 & 3.31 \\
		Encodec & 320 & 4.24 & 4.05 & 3.73 & 3.30 \\
		FunCodec & 320 & 4.29 & 4.12 & 3.86 & 3.43 \\
		FunCodec-2x & 640 & 4.29 & \textbf{4.16} & \textbf{3.94} & \textbf{3.64} \\
		FunCodec-4x  & 1280 & \textbf{4.31} & 3.94 & 3.43 & 2.91 \\
		\bottomrule
	\end{tabular}
\vspace{-0.4cm}
\end{table}
We first conduct evaluations on LibriTTS and perform architecture selection for the generalized models.
The experimental results are presented in Table \ref{tab:librtts}. 
From the table, it can be observed that our reproduced SoundStream and Encodec models achieve the-state-of-art performance at different token rates, confirming the accuracy of our implementation.
% When the log-compressed power spectrum loss is involved, the speech quality is consistently improved by FunCodec, and the improvement is more significant under lower token rates.
The inclusion of the log-compressed power spectrum loss consistently improves speech quality, particularly at lower token rates.
Building upon FunCodec, we propose low-frame-rate models with strides that are two and four times longer, resulting in FunCodec-2x and FunCodec-4x, respectively.
We find that FunCodec-2x achieves higher ViSQOL scores, while FunCodec-4x exhibits degraded performance at lower token rates.
This indicates that reducing the frame rate by two times strikes a favorable balance between time and quantization resolution. Thus, we adopt the FunCodec-2x settings for the generalized models.
% two times frame-rate reduction is a good trade-off between the time and quantization resolution.
% Therefore, we adopt the FunCodec-2x settings for generalized models.

\vspace{-0.3cm}
\subsection{Evaluation of generalized models on multiple corpus}
\begin{figure}[t!]
	% \centering
	\subfigure[Open-source models with lower token rate]{
		\centering
		\includegraphics[width=7.2cm]{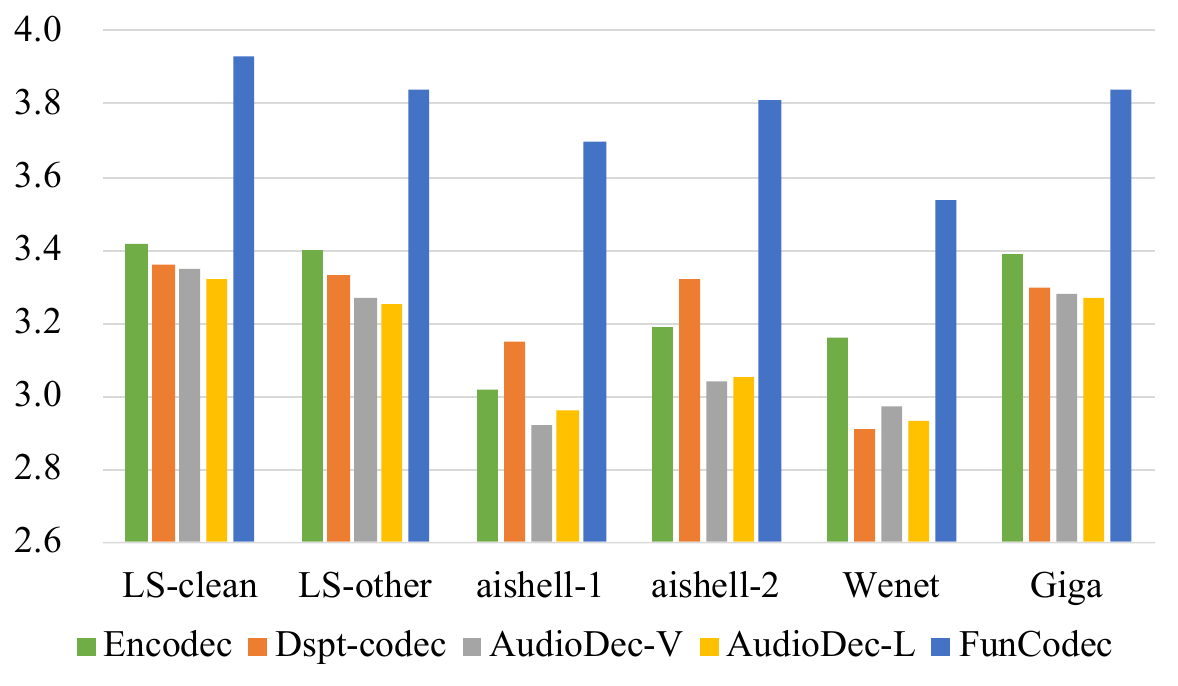}
		\label{fig:multi-corpus-lower}
	}
	\subfigure[Open-source models with higher token rate]{
		\centering
		\includegraphics[width=7.2cm]{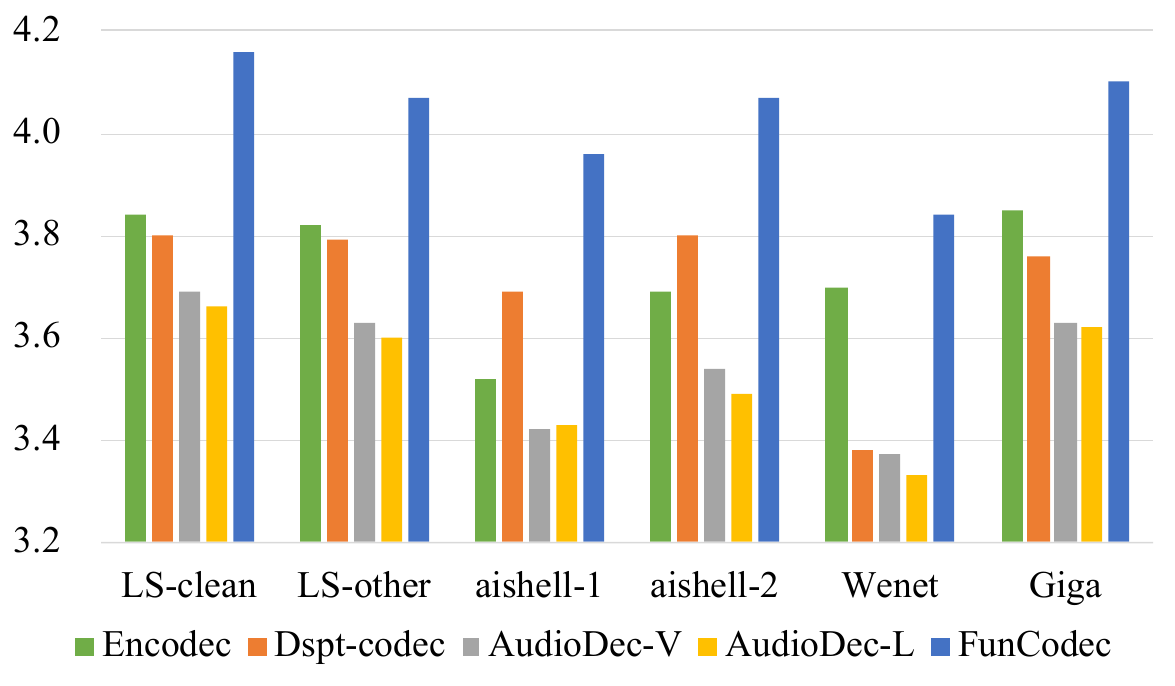}
		\label{fig:multi-corpus-higher}
	}
	\vspace{-0.3cm}
	\caption{Comparison of open-source generalized models under (a) lower and (b) higher token rate. LS denotes Librispeech test sets. While Librispeech and gigaspeech are English corpora, aishell and Wenet are Mandarin corpora.}
	\label{fig:multi-corpus-eval}
	\vspace{-0.5cm}
\end{figure}
% Open-source models can work on various sampling rates, therefore the test files are resampled to match the sampling rate.
To ensure compatibility with other open-source models, the test files are resampled to match the desired sampling rate.
Additionally, for a fair comparison, the token rate is also normalized based on the working sampling rate.
Fig. \ref{fig:multi-corpus-eval} illustrates the results of open-source models at different token rates, where lower and higher token rates correspond to the generation of 100 and 200 tokens per 16k waveform samples, respectively. 
% means the model generates 100 and 200 tokens every 16,000 waveform samples.
% We find that higher token rate always brings better compressed quality.
It was observed that higher token rates consistently yield improved compressed quality.
% Compared with other open-source models, our model can achieve much better quality for both English and Mandarin speech under the same token rate.
In comparison to other open-source models, our model demonstrates significantly better quality for both English and Mandarin speech at the same token rate.
% Surprisingly, all models perform poor on the Wenet test set. This is because the Wenet corpus is recorded in complex acoustic scenes and contains more non-speech noises. 
However, to our surprise, all models perform poorly on the Wenet test set. This can be attributed to the fact that the Wenet corpus is recorded in complex acoustic environments and contains a higher degree of non-speech noises.

\vspace{-0.2cm}
\subsection{Comparison of frequency and time domain models}
\begin{table}[t!]
	\caption{Comparison of FreqCodec and other time domain models in terms of ViSQOL score on LibriTTS. Mag denotes magnitude spectrogram. $C_{in}$ represents the channel number of inputs.}
	\vspace{0.1cm}
	\label{tab:freqcodec}
	\centering
	\setlength\tabcolsep{4.0pt}
	\begin{tabular}{l | c | c | c | c | c | c}
		\toprule
		\multirow{2}{*}{ID} & \multirow{2}{*}{Domain} & \multirow{2}{*}{Param.} & \multirow{2}{*}{Flops} & Groups & \multicolumn{2}{c}{TKR} \\
		\cline{6-7}
		&  &  &  & (Enc,Dec) & 400 & 100 \\
		\midrule
		M1 & Time & 14.85M  & 3.72G & $1,1$ & 4.29 & 3.86 \\
		M2 & Mag,Angle & 16.21M & 6.39G & $1,1$ & 4.32 & 3.84 \\
		M3 & Mag,Phase & 16.21M & 6.47G & $1,1$ & 4.36 & 3.85 \\
		M4 & Mag,Phase & 4.38M & 1.73G & $1,C_{in}$ & 4.28 & 3.79 \\
		M5 & Mag,Phase & 4.50M & 2.18G & $1,C_{in}/8$ & 4.31 & 3.81 \\
		M6 & Mag,Phase & 0.52M & 0.34G & $C_{in},C_{in}$ & 4.21 & 3.65 \\
		M7 & Mag,Phase & 0.83M & 1.03G & $C_{in}/4,C_{in}/4$ & 4.25 & 3.80 \\		
		\bottomrule
	\end{tabular}
\end{table}
Table \ref{tab:freqcodec} presents a comparison of different frequency and time domain models in terms of parameters, computation complexity and quantized speech quality. 
% Since speech signals always have clear structures on the frequency domain, 
Due to the distinct structures in the frequency domain, FreqCodec models demonstrate superior speech quality at higher token rates. 
% Compared with the representation of magnitude and angle spectra, the representation of magnitude and normalized phase spectra is more appropriate for speech codec (seen in M2 and M3). 
Notably, the representation of magnitude and normalized phase spectra, as observed in M2 and M3, proves to be more suitable for speech signals compared to the magnitude and angle spectra. 
% In addition, we further reduce the number of parameters and computation complexity using depthwise convolutions with different group settings. 
Additionally, we successfully reduce the number of parameters and computational complexity through the utilization of depthwise convolutions.
% The results of M4-M7, we find that, with proper group split on encoder and decoder, the model and computation complexity can be much reduced with negligible quality degradation.
The results of M4-M7 indicate that, by appropriately splitting the groups in the encoder and decoder, both the model and computational complexity can be significantly reduced without compromising quality.

\subsection{The impact of semantic augmentation}
In contrast to other audio signals, speech carries explicit semantic information. 
As a result, we enhance our codec models by incorporating force-aligned phoneme labels.
Results in Table \ref{tab:mfa-ppg} demonstrate that the inclusion of semantic tokens consistently enhances the quality of quantized speech.
%  we find that involving semantic tokens can consistently improve the quality of quantized speech. 
Moreover, the ``residual'' combination method appears to be more suitable for speech codecs. 
% This finding reveals that decoupling the semantic and acoustic information is a reasonable approach to reduce the token rate.
This finding highlights the viability of decoupling semantic and acoustic information as an effective approach for reducing token rates.

\begin{table}[t!]
	\vspace{-0.4cm}
	\caption{The impact of semantic information under low token rates.}
	\vspace{0.1cm}
	\label{tab:mfa-ppg}
	\centering
	\setlength\tabcolsep{10.0pt}
	\begin{tabular}{l | c | c | c}
		\toprule
		ID & Info. Combiner & TKR=100 & TKR=50 \\
		 \midrule
		 M1 & None & 3.86 & 3.43 \\
		 SM1 & Concat. & 3.87 & 3.69 \\
		 SM2 & Residual & \textbf{3.93} & \textbf{3.60} \\
		 SM3 & Addition & 3.92 & 3.59 \\
		\bottomrule
	\end{tabular}
\end{table}

\vspace{-0.3cm}
\subsection{Applying to down-stream tasks}
\begin{table}[t!]
	\vspace{-0.4cm}
	\caption{Comparison of different inputs for ASR on Librispeech.}
	\vspace{0.1cm}
	\label{tab:codec-asr}
	\centering
	\setlength\tabcolsep{3.5pt}
	\begin{tabular}{l c c c c}
		\toprule
		Inputs & dev-clean & dev-other & test-clean & test-other \\
		\midrule
		fBank & 2.68 & 7.31 & 3.01 & 7.00 \\
		\midrule
		Codec Embedding & 4.83 & 13.80 & 4.85 & 14.49 \\
		\ \ with SpecAug & 3.29 & 9.57 & 3.59 & 9.67 \\
		Codec Index & 6.98 & 19.06 & 7.00 & 19.84 \\
		\ \ with SpecAug & 4.37 & 12.43 & 4.48 & 12.81 \\
		\bottomrule
	\end{tabular}
\vspace{-0.4cm}
\end{table}
Except for speech quality, we also evaluate FunCodec in downstream tasks. 
Table \ref{tab:codec-asr} presents the results of ASR task (TKR=100). Synthesis examples of VALL-E TTS system can be found on Demo pages.
We find that quantized tokens preserve a large proportion of speech content, leading to a low recognition error rate.
% Compared with the codec index, the embedding of codebook is also important for ASR task.
Furthermore, the codebook embedding emerges as a crucial factor for the ASR task.
By comparing the results of clean and other tests, it is observed that codec-based discrete inputs are more sensitive to the acoustic environment than continuous fbank features.

\vspace{-0.25cm}
\section{Summary}
\vspace{-0.25cm}
In this paper, we introduce the open-source toolkit, FunCodec, which is reproducible and integrable. 
% The release of the codebase is accompanied by pre-trained academic and generalized models, which are available through ModelScope.
Along with the codebase, pre-trained academic and generalized models are released on Huggingface and ModelScope. 
Based on FunCodec, we further evaluate the frequency domain models and augment the RVQ module with semantic information.
In addition, we also validate FunCodec models in the context of down-stream tasks, including ASR and TTS.
% Experimental results show that FunCodec implements state-of-the-art codec models and 

\bibliographystyle{IEEEbib}
\bibliography{refs}

\end{document}